\documentclass[fleqn,10pt, twocolumn]{wlscirep}

\usepackage[table]{xcolor}
\usepackage[utf8]{inputenc}
\usepackage[T1]{fontenc}

\usepackage{amsmath}
\usepackage{amssymb}
\usepackage{glossaries}
\usepackage{graphicx}
\usepackage{here}
\usepackage{times}
\usepackage{float}
\usepackage{booktabs}
\usepackage{bm}
\usepackage{siunitx}
\usepackage{multirow}
\usepackage{mathtools}
\usepackage{xspace}
\usepackage[none]{hyphenat} 

\usepackage{makecell}
\usepackage{tabularx}
\usepackage{colortbl}

\title{Defects and Structural Evolution in CrSBr: Machine-Learning Potentials for Atomistic Simulations}

\author[1,2]{Arsalan Hashemi\thanks{Corresponding author: arsalan.hashemi@uef.fi}}
\author[3]{Daniil Kruklinskii}
\author[4]{Xiaoya Chang}
\author[4,5]{Tapio Ala-Nissila}
\author[1,2]{Mikko Karttunen}
\author[3]{Arkady V. Krasheninnikov}
\author[3]{Mahdi Ghorbani-Asl\thanks{Corresponding author: mahdi.ghorbani@hzdr.de}}

\affil[1]{European Laboratory for Learning and Intelligent Systems (ELLIS)  Institute Finland, Maarintie 8, 02150 Espoo, Finland}
\affil[2]{Department of Technical Physics, University of Eastern Finland, P.O. Box 1627, FI-70211 Kuopio, Finland}
\affil[3] {Institute of Ion Beam Physics and Materials Research, Helmholtz-Zentrum Dresden-Rossendorf, 01328 Dresden, Germany}
\affil[4]{MSP Group, Department of Applied Physics, Aalto University, P.O. Box 15600, FI-00076 Aalto, Espoo, Finland}
\affil[5]{Interdisciplinary Centre for Mathematical Modelling and Department of Mathematical Sciences, Loughborough University, Loughborough, Leicestershire LE11 3TU, United Kingdom}

\begin{abstract}
The layered van der Waals semiconductor CrSBr combines intrinsic magnetic order with semiconducting and optical properties, making it a promising material for multifunctional electronic and photonic applications. Understanding how defects and structural transformations affect its properties under experimentally relevant conditions, however, remains challenging for first-principles simulations due to the relevant length and time scales. Here, we develop and benchmark machine-learning interatomic potentials (MLIPs) for the Cr--S--Br system, enabling large-scale atomistic simulations of CrSBr with accuracy approaching that of density functional theory (DFT). The charge-aware variant qNEP achieves near-DFT accuracy in predicting lattice constants, formation enthalpies, vacancy energetics, and the anisotropic pressure response of $o$-CrSBr. Nudged elastic band calculations reveal that pre-existing S vacancies lower the Cr Frenkel-pair migration barrier by up to 1.1 eV, providing an atomistic mechanism for beam-induced Cr redistribution and structural transformations. Displacement threshold calculations further show that Br is the most susceptible species to electron-beam-induced displacement, identifying the relative ease of removing different atomic species. Together, these results establish a validated computational framework for investigating atomic-scale disorder and structural transformations in CrSBr, providing a basis for defect engineering in materials for future magneto-optoelectronic applications.

\end{abstract}

\begin{document}

\flushbottom
\maketitle
%

\section*{Introduction}
Chromium sulfur bromide (CrSBr) is a van der Waals (vdW) layered magnetic semiconductor that is amenable to mechanical exfoliation into thin magnetic flakes~\cite{Ziebel2024_nl,klein2024materials,GOSER1990129}.
The material exhibits a direct bandgap in the visible spectral range and crystallizes in the \emph{orthorhombic} phase ($o$-CrSBr), wherein monolayers are antiferromagnetically coupled to neighboring layers along the crystallographic stacking direction.
This combination of electronic structure and magnetic order has motivated substantial interest in integrating CrSBr into spintronic and magneto-optoelectronic device concepts, including magneto-optoelectronic switching elements~\cite{wilson2021interlayer,telford2022coupling}, excitonic information carriers~\cite{bae2022exciton}, photothermoelectric photodetectors~\cite{Zhou2025_acsphotonics}, and magnetic proximity-effect heterostructure layers~\cite{yang2024electrostatically,jo2023local}.

CrSBr is further distinguished by its propensity for stimulus-induced local phase transformations, which can be initiated by electron/ion-beam irradiation~\cite{klein2022control,Deng2023_aem} or by the application of external pressure~\cite{langston2025pressure}.
In both scenarios, the ambient-stable $o$-CrSBr is converted into a metastable \emph{monoclinic} polymorph ($m$-CrSBr), accompanied by a reorientation of the interlayer stacking.
Notably, irradiation-driven experiments indicate that this transformation proceeds via an intermediate metastable configuration involving Cr migration into the vdW gap.
By contrast, under hydrostatic pressure, the structural pathway and mechanistic determinants governing the phase transition remain insufficiently resolved.

The picture becomes richer still when defects are considered, even in their most elementary form, i.e., intrinsic point defects.
Such defects are thermodynamically unavoidable at any nonzero temperature~\cite{Ashcroft76}; in atomically thin two-dimensional (2D) materials, however, their role extends well beyond a mere thermodynamic inevitability.
The reduced atomic coordination and the absence of a bulk environment to screen local perturbations render point defects strong determinants of the electronic, magnetic, and optical properties~\cite{krasheninnikov2007engineering,wang2012electronics}.
This sensitivity opens a direct route to property engineering: controlled introduction of defects provides a means to modulate defect populations, enabling targeted tuning of material functionalities~\cite{Long2024_APR}.

Prior studies have established that $o$-CrSBr routinely harbors high concentrations of intrinsic point defects, with bromine vacancies ($\mathrm{V}_{\mathrm{Br}}$) dominating at areal densities around $10^{13}~\text{cm}^{-2}$, while Cr and S vacancies occur roughly two orders of magnitude less frequently~\cite{Torres2023_AFM}.
Beyond simple vacancies, Cr Frenkel pairs, comprising a Cr vacancy ($\mathrm{V}_{\mathrm{Cr}}$) and a Cr interstitial ($\mathrm{Cr}_{\mathrm{i}}$) that preferentially stabilizes within the vdW gap, have been identified as a distinct defect motif~\cite{Long2023_NL}.
More complex assemblies, including $\mathrm{V}_{\mathrm{Br-Cr-Br}}$, $\mathrm{V}_{\mathrm{Br-Cr}}$, $\mathrm{V}_{\mathrm{2Br}}$, and $\mathrm{V}_{\mathrm{2Cr}}$, have been cataloged by electron microscopy and deep-learning-assisted analysis of bilayer CrSBr~\cite{PhysRevX.15.021080,Klein_acs_nano_2023}.
Surface-sensitive measurements further indicate an elevated density of $\mathrm{V}_{\mathrm{Br}}$ at crystal surfaces and a reduced density of $\mathrm{V}_{\mathrm{Cr-S}}$ in the bulk~\cite{Klein_acs_nano_2023}, while controlled He$^{+}$ irradiation demonstrates that defect populations can be tuned systematically with beam energy~\cite{Torres2023_AFM}.

Despite these advances, it remains unclear how defect complexes form and how stable they are.
The atomic-scale mechanisms governing Cr redistribution during irradiation and annealing are also poorly understood.
Addressing them requires simulations at length and time scales that are inaccessible to density functional theory (DFT), whose computational cost becomes prohibitive for the large supercells and extensive configurational sampling that finite-temperature defect physics demands.
Machine-learning interatomic potentials (MLIPs) offer a direct route out of this bottleneck.
By fitting a surrogate model to DFT reference energies, forces, and virials, MLIPs reproduce the potential energy surface at a fraction of the computational cost, enabling large-scale molecular dynamics (MD) simulations and systematic defect searches that would otherwise be impractical~\cite{Leimeroth_2025_iop,JACOBS2025101214}.
Unlike empirical force fields, they require no assumptions about functional form and can, in principle, be made as accurate as the underlying DFT reference.

An accurate and transferable MLIP for CrSBr has yet to be developed, and constructing such a model presents several challenges.
The model must describe a broad configurational space encompassing layered and defective structures, competing orthorhombic and monoclinic polymorphs, and irradiation-induced amorphous intermediates.
Magnetic interactions introduce additional complexity because the potential energy surface depends on the spin configuration.
The small energy splitting between ferromagnetic and antiferromagnetic ordering in pristine $o$-CrSBr ($\approx 0.06$~meV/atom) motivates a geometry-based description of its structural energetics, although this observation alone does not establish its validity for defects or structural transformations.
Careful validation across these configurations is therefore essential for reliable simulations of defect dynamics and finite-temperature structural evolution.

Here, we present a curated Cr--S--Br dataset and develop MLIPs to investigate defects under strain, at finite temperatures, and under electron-beam irradiation.
This work provides a basis for the predictive engineering of defect complexes in CrSBr and supports the development of air-stable 2D magnets for magneto-optoelectronic applications.

\section*{Computational Framework}
\label{sec:computational_detail}
\subsection*{Density Functional Theory Calculations}
\label{section:dft}

All DFT calculations were carried out using the Vienna \textit{Ab initio} Simulation Package (VASP)~\cite{vasp1,vasp2} within the projector augmented-wave (PAW) framework~\cite{paw}.
For chromium, the \texttt{Cr\_pv} PAW dataset was employed, thereby treating the Cr semicore $3p$ states explicitly as valence.
This selection is consistent with the VASP PAW-potential recommendations for $3d$ transition metals and is expected to enhance transferability for locally distorted coordination environments, albeit at the expense of an increased plane-wave basis requirement.
Exchange--correlation effects were described using the Perdew--Burke--Ernzerhof (PBE) generalized-gradient approximation~\cite{pbe}.
Long-range dispersion interactions were incorporated via the Grimme DFT-D3 correction~\cite{grimmed3} to capture interlayer vdW contributions.
Electronic self-consistency iterations were converged until the change in total energy between successive electronic steps was less than $10^{-6}$~eV.
When required, structural optimization was iterated until the maximum residual interatomic force decreased below $10^{-2}$~eV/\AA.
A plane-wave kinetic-energy cutoff of 500~eV was used.
Brillouin-zone sampling was performed with a $\mathbf{k}$-point mesh corresponding to a reciprocal-space density of 0.25~\AA$^{-1}$.
Spin polarization was included throughout.

The magnetic configuration was initialized with antiferromagnetic interlayer coupling, while each individual monolayer was ferromagnetic.
Initial magnetic moments were assigned using a structure-informed protocol: S and Br sites were initialized with zero magnetic moment, whereas Cr sites were initialized with alternating moments of $+3$ and $-3~\mu_\text{B}$ according to the monolayer ordering along the $c$-axis.
Magnetic layers were identified from the atomic coordinates by projecting Cr positions onto the stacking direction, explicitly accounting for periodic boundary conditions, and merging closely spaced Cr sub-layers into a single effective monolayer, as commonly encountered in more amorphous configurations.
The same constraints were applied to all elemental, binary, and ternary structures in the Cr--S--Br dataset.

Ab initio molecular dynamics simulations were carried out to investigate atomic displacements caused by electron collisions. A time step of 0.5~fs and a single $\Gamma$-point sampling of the Brillouin zone were used. The simulations were performed using a $5 \times 5$ supercell containing 50 atoms of each element (Cr, S, and Br).


\subsection*{Reference Dataset}
\label{section:dataset}

\begin{figure*}[!t]
    \centering
    \includegraphics[width=\textwidth]{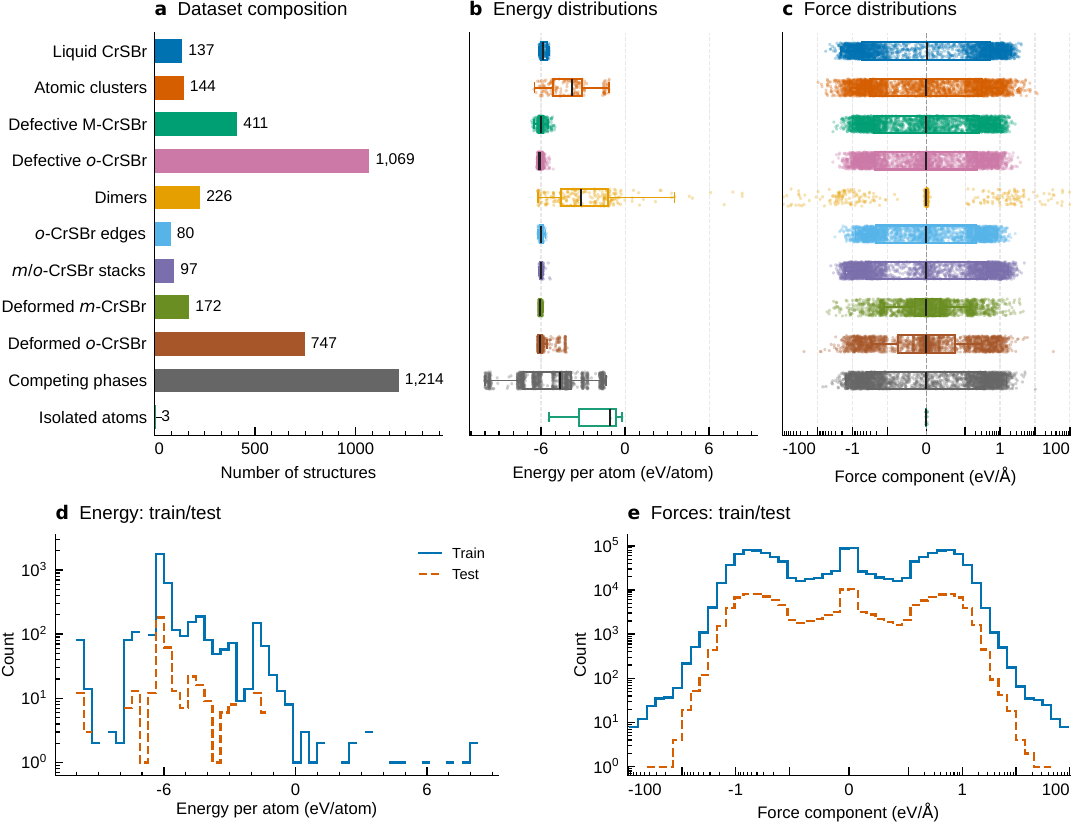}
    \caption{\textbf{Overview of the Cr--S--Br reference dataset.}
    \textbf{a}, Number of structures per subset.
    \textbf{b} and \textbf{c}, Distribution of DFT energies per atom and Cartesian force components for each subset; each point represents one configuration.
    Boxes show the interquartile range, central lines mark the medians, and whiskers extend to the furthest data points within 1.5 interquartile ranges of the box.
    \textbf{d} and \textbf{e}, Counts of energy and force in the training and test sets.}
    \label{fig:dataset}
\end{figure*}
The reference dataset was generated through an iterative active-learning loop.
An initial MLIP model was trained on a seed set of isolated atoms, dimers, mono- and heteroatomic clusters, and strained, rattled $o$–CrSBr structures.
The resulting model was used for MD simulations and then updated after DFT recomputation of configurations identified as physically questionable or structurally new.
This closed-loop procedure was iterated, progressively expanding the dataset, until the MD trajectories remained dynamically stable under all thermodynamic conditions of interest and the full relevant configurational space was covered.
Details of the stepwise data expansion are provided in the Supplementary Information (SI), Section S1.

Figure~\ref{fig:dataset}(a) presents data subsets, including isolated C, S, and Br atoms; distinct dimers; atomic clusters; a distorted pristine monolayer-to-bulk CrSBr phases; $m$/$o$-CrSBr heterostructures; $o$-CrSBr nanoribbon edges; and a range of intrinsic defects at different concentrations spanning monolayer, multilayer, and bulk $m$- and $o$-CrSBr, as well as liquid CrSBr generated via high-temperature MD amorphization simulations.
Point-defect configurations constitute the largest structural class: 1,069 orthorhombic and 411 monoclinic structures collectively cover Br, S, and Cr vacancies at a range of concentrations and supercell sizes, self-interstitials, antisites, Frenkel pairs, interlayer bridging atoms, sampled Cr migration intermediates, and mixed ensembles of coexisting point defects.
Note that recrystallization, for example, under irradiation requires the involvement of different Cr/S/Br phases~\cite{Deng2023_aem}.
To ensure transferability across competing crystalline environments, we also incorporated 1,214 structures taken from elemental polymorphs of Br, Cr, and S and from binary compounds (SBr, CrBr$_2$, CrBr$_3$, Cr$_5$S$_6$, Cr$_3$S$_4$, Cr$_2$S$_5$, Cr$_2$S$_3$, CrS$_2$, and CrS).
In total, 20 distinct crystalline structures, including stable and metastable polymorphs, were added to the dataset.
These were obtained from the Materials Project database~\cite{Jain2013_mat_projects} and further enriched through MD sampling.

The final dataset includes 4,300 labeled configurations with energy, force, and virial annotations.
Isolated atoms, dimers, and clusters (373 data points) were used only for training.
The remaining structures were randomly shuffled and partitioned (90/10) into training and test sets containing 3,908 and 392 structures, respectively.
Overall, the dataset contains per atom energies ranging from approximately $-10$ to $9$ eV and force components spanning multiple orders of magnitude (Fig.~\ref{fig:dataset}(b--c)).
This wide range captures the diversity of chemical environments necessary to achieve a transferable description of CrSBr under defect-induced, thermal, and mechanical perturbations.
Configurations dominated by strong short-range repulsion (e.g., dimers) give rise to large force magnitudes and positive energies; therefore, these high-energy/high-force regimes were intentionally represented in the training set rather than being shared with the test set (Fig.~\ref{fig:dataset}(d--e)).

\subsection*{Machine-Learning Interatomic Potentials}
\label{section:mlips}

Over the last few years, a wide range of different types of MLIPs have been developed~\cite{Leimeroth_2025_iop,JACOBS2025101214}, yet it is important to realize which approach is the best for a given problem.
For the migration pathways considered here, MACE~\cite{Batatia2022MACE} may reproduce DFT energy barriers more accurately, whereas NEP~\cite{Zheyong_2021_PRB} may underestimate them.
Nevertheless, NEP's computational efficiency enables larger systems and longer trajectories, supporting extensive exploration of irradiation-induced defects and subsequent annealing~\cite{LiuPRB_2023}.
Underestimated barriers may make atomic rearrangements occur more readily during annealing.
However, they can also alter which pathways are favored and how often particular defects form, so this speedup should not be interpreted as validated accelerated dynamics~\cite{Voter1997}. NEP can nevertheless provide an efficient way to identify candidate defect structures, which can then be assessed alongside their associated energy barriers using DFT.

All MLIPs developed in this work were trained using the NEP framework implemented in the open-source GPUMD package~\cite{Fan2017}.
In NEP, the total potential energy is expressed as a sum of atomic energies, each predicted by a feed-forward neural network from descriptors of the atom's local chemical environment.
The network weights are optimized using the separable natural evolution strategy (SNES), an evolutionary algorithm that trains directly on GPU hardware.
During dataset refinement, model retraining was accelerated using TorchNEP~\cite{wu2026torchnepultraefficientaccuratetraining}, a PyTorch reimplementation of the NEP architecture that replaces the evolutionary optimizer with analytically derived gradients and adaptive optimization within a two-stage training protocol.
TorchNEP reduces training time by more than two orders of magnitude relative to the original SNES-based  .
To account for long-range electrostatic interactions, we also used qNEP, a charge-aware extension of the NEP framework~\cite{10.1021/acs.jctc.6c00146}.
In qNEP, each atom is assigned an environment-dependent partial charge predicted by an auxiliary neural network, enabling a physically consistent treatment of long-range electrostatic interactions.

The radial and angular cutoffs were set to 5.2 and 6.0~\AA, respectively, and the neural network contained 50 neurons.
The \texttt{basis\_size} and \texttt{n\_max} parameters were set to $(8,8)$ and $(9,7)$, respectively.
A Ziegler–Biersack–Littmark (ZBL)\cite{ZBL} repulsive potential with a cutoff of 1.2~\AA\ was incorporated during MLIP training to accurately describe short-range atomic interactions during energetic collisions and displacement processes.
NEP and qNEP were trained for $4\times10^{5}$ generations.
TorchNEP was trained for 1,500 epochs, with an initial learning rate of $1\times10^{-3}$ and a second stage beginning at epoch 700 with a learning rate of $5\times10^{-4}$.

%
\subsection*{Molecular Dynamics and Barrier Simulations}
\label{section:md}

MD simulations were performed in either the $NVT$ or $NPT$ ensemble within the Berendsen thermostat~\cite{Berendsen_JCP_1984}.
To expand the dataset, 100 ps MD simulations were performed with a timestep of 2 fs across a range of temperatures and pressures.
Longer simulations under elevated-temperature and high-pressure conditions were used to assess the stability of the potentials.
For simulations of pressure-induced lattice changes, an $NPT$ ensemble was used with 100 ps of equilibration followed by a 200 ps production run at 300 K.
Structural relaxations were performed using the FIRE algorithm with an interatomic force tolerance of $10^{-6}$ eV/\AA.

The minimum-energy pathways were determined using the climbing-image nudged elastic band (CI-NEB) method, as implemented in the Atomic Simulation Environment (ASE) Python library~\cite{ASE}.
Each pathway comprised 15 images, including the two endpoints.
The intermediate images were initialized using image-dependent pair-potential (IDPP) interpolation with the minimum-image convention.
The improved-tangent formulation was employed with a spring constant of 0.20 eV/Å$^2$.
The band was optimized using the FIRE algorithm through successive regular- and climbing-image stages with progressively tighter convergence criteria, until the maximum NEB force fell below 0.05 eV/\AA.
The MLIPs were interfaced with ASE through the CALORINE~\cite{CALORINE} package.

\section*{Results}
\label{sec:results}

\subsection*{Validation of NEP variants}
\label{subsec:mlvsdft}

\begin{figure*}[!t]
    \centering
    \includegraphics[width=\textwidth]{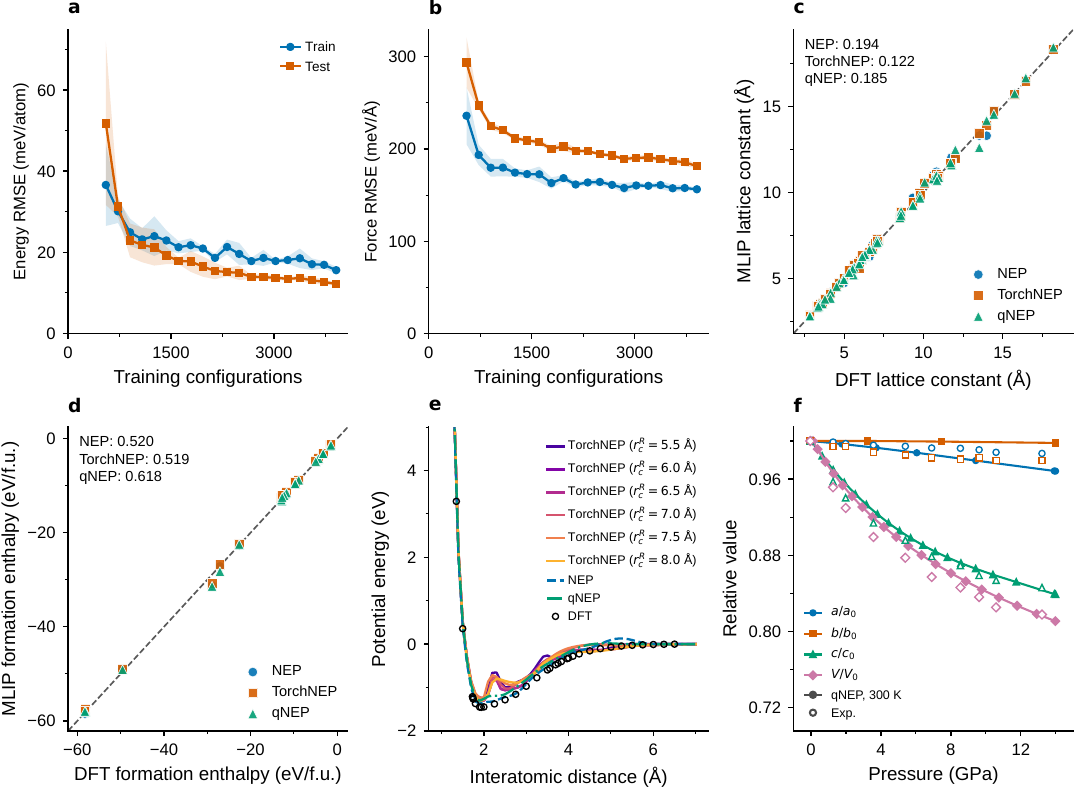}
    \caption{\textbf{Learning behavior and validation of the MLIPs.}
    \textbf{a,b,} Energy and force root-mean-square errors (RMSEs) for the training and test sets as a function of the number of training configurations. Shaded regions indicate one standard deviation across five independent data splits and training runs.
    \textbf{c,d,} Lattice constants and formation enthalpies per formula unit predicted by NEP, TorchNEP and qNEP compared with DFT values. Dashed lines indicate perfect agreement; annotations report the RMSE for each MLIP.
    \textbf{e,} Cr--Cr dimer potential-energy curves calculated using DFT, NEP, qNEP and TorchNEP with different cutoff radii.
    Each curve is shifted to zero at its largest sampled separation.
    \textbf{f,} Pressure dependence of the normalized lattice parameters ($a/a_0$, $b/b_0$, and $c/c_0$) and volume ($V/V_0$), comparing qNEP simulations at 300~K (lines and filled markers) with experimental data (open markers; extracted from Ref. [\citenum{q4f9-4bvz}]). Each series is normalized by its own value at the lowest sampled pressure.
    }
    \label{fig:mlip_validation}
\end{figure*}
The predictive accuracy of the NEP variants was assessed using learning curves, equilibrium lattice constants, formation enthalpies, Cr--Cr dimerization energies, and the pressure-dependent structural response of $o$-CrSBr, as summarized in Fig.~\ref{fig:mlip_validation}.
The learning curves were generated using TorchNEP owing to its fast training.
Given the common NEP framework, these curves may also offer qualitative guidance on the training-data requirements of the other variants.
The energy and force RMSEs (Figs.~\ref{fig:mlip_validation}(a,b)) decrease substantially with increasing training-set size, with more gradual improvements beyond approximately 1500 configurations.
For the largest training set, the test RMSEs are 13~meV/atom for energies and 180~meV/\AA\ for forces, compared with training RMSEs of 16~meV/atom and 155~meV/\AA{}, respectively.
The lower test energy error mainly reflects differences in dataset composition: isolated atoms, dimers, and atomic clusters were included only in the training set and contributed substantially to its energy error.
Finally, TorchNEP was trained on the full dataset, yielding training RMSEs of $\approx$12~meV/atom for energies and 154~meV/\AA\ for forces.
Training on the full dataset yielded energy and force RMSEs of 16~meV/atom and 195~meV/\AA\ for NEP, and 17~meV/atom and 192~meV/\AA\ for qNEP, respectively.

All three models reproduce the DFT lattice constants of the investigated competing phases, including ternary compounds (Fig.~\ref{fig:mlip_validation}(c)).
TorchNEP achieves the lowest RMSE of 0.122~\AA, compared with 0.194 and 0.185~\AA\ for NEP and qNEP, respectively.
The formation enthalpies also follow the DFT reference values over a broad energy range (Figs.~\ref{fig:mlip_validation}(c)), with RMSEs of 0.520, 0.519, and 0.618~eV per formula unit for NEP, TorchNEP, and qNEP, respectively.
Thus, TorchNEP provides the closest agreement for lattice constants, while NEP and TorchNEP exhibit comparable accuracy for formation enthalpies.
The corresponding values are summarized in Table~S1.

The Cr--Cr dimer curves provide a complementary assessment of the interaction energy as a function of interatomic separation (Fig.~\ref{fig:mlip_validation}(e)).
The models capture the steep short-range repulsion and the attractive well near 2~\AA, but exhibit deviations from DFT at intermediate separations.
Varying the TorchNEP cutoff between 5.5 and 8.0~\AA\ changes the shape of this region without yielding a uniformly improved description.
These differences highlight the sensitivity of low-coordination interactions to the model parametrization.
NEP and qNEP reproduce the potential well more accurately than TorchNEP, although all three models underestimate the binding strength relative to DFT.
NEP also exhibits a spurious repulsive contribution near the dissociation limit, whereas qNEP provides the most reliable overall description of the Cr--Cr interaction.
The corresponding dimerization energies and structural data for the other dimers are reported in Table~S2 and Fig.~S1; for these dimers, all three models show close agreement with DFT.
Based on this comparative assessment, qNEP was selected for subsequent calculations, with the other variants included where relevant for comparison.

Finally, qNEP-MD simulations at 300~K reproduce the strongly anisotropic compression observed experimentally for $o$-CrSBr~\cite{q4f9-4bvz} (Fig.~\ref{fig:mlip_validation}(f)).
The $c$ lattice parameter contracts substantially with increasing pressure, whereas $a$ and $b$ change comparatively little.
The calculated volume decreases accordingly, although quantitative deviations from experiment become more apparent at higher pressures.
A comparison with other models is shown in Fig.~S2.
Together, these benchmarks establish the accuracy of the models for the tested equilibrium properties while identifying remaining limitations in dimer energetics and the pressure response.

\subsection*{Pristine CrSBr phases}
\label{sec:lattice-constant}

To further assess the structural accuracy of MLIPs, we first compare the calculated lattice constants of pristine CrSBr with DFT and experimental results.
For the $o$-CrSBr, DFT yields $a=3.512$~\AA\ and $b=4.692$~\AA, whereas qNEP gives $a=3.525$~\AA\ and $b=4.738$~\AA, which are in good agreement with the experimental values of $a=3.51$~\AA and $b=4.77$~\AA~\cite{klein2022control,Deng2023_aem}.
Accounting for the doubled periodicity along $c$ in the computational cell, the DFT and qNEP values correspond to $c/2=7.876$ and $7.874$~\AA, respectively, within 0.5\% of the experimental value of 7.907~\AA~\cite{Deng2023_aem}.
For $m$-CrSBr, qNEP predicts $(a,b,c)=(6.112,7.188,11.733)$~\AA, closely matching the DFT values of $(6.098,7.170,11.705)$~\AA.
Furthermore, qNEP reproduces the DFT energetic preference for the orthorhombic phase, placing the monoclinic phase 0.209~eV per formula unit higher in energy, compared with 0.202~eV from DFT.

\begin{table}[!t]
\centering
\small
\setlength{\tabcolsep}{9pt}
\renewcommand{\arraystretch}{1.12}
\caption{\textbf{Defect formation energies.}
Formation energies (eV) are reported for bilayer vacancies and monolayer antisites in $o$- and $m$-CrSBr.
RMSEs quantify the qNEP deviations from DFT within each group.
}
\label{tab:defect_formation_qnep}

\begin{tabular}{@{}l
    *{2}{S[table-format=1.3]}
    @{\hspace{2mm}}
    l *{2}{S[table-format=1.3]}@{}}
\toprule
\multicolumn{3}{c}{\textbf{Bilayer vacancies}} &
\multicolumn{3}{c}{\textbf{Monolayer antisites}} \\
\cmidrule(lr){1-3}\cmidrule(lr){4-6}
Defect & {DFT} & {qNEP} &
Defect & {DFT} & {qNEP} \\

\midrule
\multicolumn{6}{c}{$o$-CrSBr} \\
\addlinespace[3pt]

V$_{\mathrm{Br}}^{\mathrm{in}}$ & 2.195 & 2.285 &
$\mathrm{Br}_{\mathrm{S}}$      & 0.706 & 0.701 \\

V$_{\mathrm{Br}}^{\mathrm{out}}$ & 2.067 & 2.040 &
$\mathrm{Br}_{\mathrm{Cr}}$      & 3.696 & 3.694 \\

V$_{\mathrm{S}}^{\mathrm{in}}$ & 2.855 & 2.692 &
$\mathrm{S}_{\mathrm{Cr}}$     & 3.625 & 4.230 \\

V$_{\mathrm{S}}^{\mathrm{out}}$ & 2.822 & 2.628 &
$\mathrm{S}_{\mathrm{Br}}$      & 1.846 & 0.910 \\

V$_{\mathrm{Cr}}^{\mathrm{in}}$ & 3.194 & 3.119 &
$\mathrm{Cr}_{\mathrm{Br}}$     & 4.416 & 4.815 \\

V$_{\mathrm{Cr}}^{\mathrm{out}}$ & 3.186 & 3.104 &
$\mathrm{Cr}_{\mathrm{S}}$       & 3.752 & 4.161 \\

V$_{\mathrm{Cr}}\!-\!\text{V}_{\mathrm{Br}}$
    & 4.103 & 4.159 & & & \\

($\text{V}_{\mathrm{Cr}}\!-\!\mathrm{Cr}_{\mathrm{i}}$)$_{a}$
    & 3.445 & 3.500 & & & \\

($\text{V}_{\mathrm{Cr}}\!-\!\mathrm{Cr}_{\mathrm{i}}$)$_{b}$
    & 2.423 & 2.428 & & & \\

\addlinespace[3pt]
RMSE & & 0.101 & RMSE & & 0.511 \\

\midrule
\multicolumn{6}{c}{\textit{$m$-CrSBr}} \\
\addlinespace[3pt]

V$_{\mathrm{Br}}^{\mathrm{in}}$ & 2.608 & 2.551 &
$\mathrm{Br}_{\mathrm{S}}$      & 0.342 & 0.161 \\

V$_{\mathrm{Br}}^{\mathrm{out}}$ & 2.518 & 2.395 &
$\mathrm{S}_{\mathrm{Br}}$       & 0.345 & 0.259 \\

V$_{\mathrm{S}}^{\mathrm{in}}$ & 2.649 & 2.579 &
$\mathrm{S}_{\mathrm{Cr}}$     & 2.383 & 2.296 \\

V$_{\mathrm{S}}^{\mathrm{out}}$ & 2.541 & 2.435 &
$\mathrm{Cr}_{\mathrm{Br}}$     & 4.651 & 4.151 \\

V$_{\mathrm{Cr}}$        & 2.005 & 2.080 &
$\mathrm{Cr}_{\mathrm{S}}$ & 5.155 & 4.601 \\

V$_{\mathrm{Cr}}\!-\!\text{V}_{\mathrm{S}}$
    & 3.004 & 3.145 & & & \\

V$_{\mathrm{Cr}}\!-\!\text{V}_{\mathrm{Br}}$
    & 3.964 & 3.961 & & & \\

\addlinespace[3pt]
RMSE & & 0.093 & RMSE & & 0.348 \\

\bottomrule
\end{tabular}
\end{table}
\subsection*{Defect formation energies}
\label{sec:defetc_formation}

Defect formation energies provide a test of the ability of qNEP to describe changes in local coordination and chemical composition beyond pristine structures.
Table~\ref{tab:defect_formation_qnep} compares DFT and qNEP results for vacancies and vacancy-containing complexes in bilayer CrSBr, together with antisite defects in monolayers of the orthorhombic and monoclinic phases.
The symbol V$_X$ indicates that atom $X$ is absent, while $X_Y$ refers to an atom of type $X$ located on a lattice site normally occupied by species $Y$.
For $o$-CrSBr, a $5 \times 4 \times 1$ supercell was selected, whereas for $m$-CrSBr a $3 \times 3 \times 1$ supercell was employed as a platform for defect modeling.
A vacuum of 25~\AA\ was introduced perpendicular to the atomic planes in all structures.

A monolayer of $o$-CrSBr comprises a puckered Cr--S framework terminated by Br atoms on both surfaces.
The framework consists of two coupled, buckled Cr--S sublayers, with each Cr atom coordinated by four S atoms and two Br atoms in a distorted octahedral environment.
These CrS$_4$Br$_2$ octahedra share S--Br edges along the $a$ direction, S corners along the $b$ direction, and S--S edges along the in-plane diagonals, forming an anisotropic rectangular network.
Therefore, in bilayer $o$-CrSBr, each monolayer contains two atomic sublayers for each element.
We label the sites closer to the interlayer gap as ``in'' and those closer to the outer surfaces of the bilayer as ``out''.
Cr Frenkel pairs are formed by displacing an inner Cr atom into the vdW gap, with a lateral displacement along either the $a$- or $b$-axis, yielding the ($\text{V}_{\mathrm{Cr}}\!-\!\mathrm{Cr}_{\mathrm{i}}$)$_{a}$ and ($\text{V}_{\mathrm{Cr}}\!-\!\mathrm{Cr}_{\mathrm{i}}$)$_{b}$ defect complexes, respectively.

\begin{figure*}[!t]
    \centering
    \includegraphics[width=\textwidth]{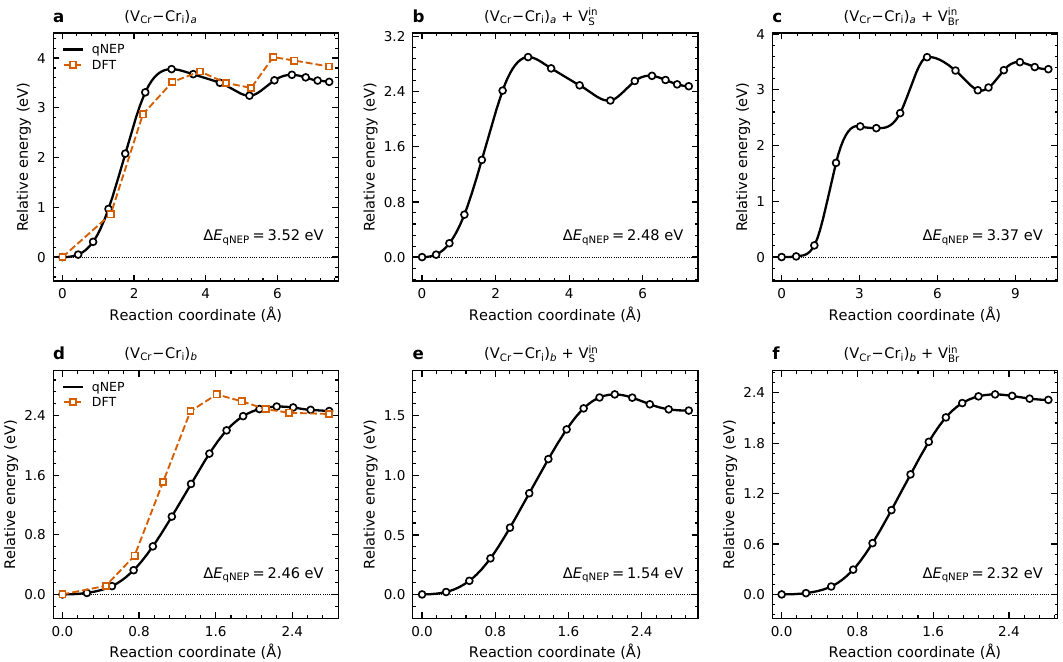}
    \caption{\textbf{Cr Frenkel-pair formation pathways in bilayer CrSBr.}
    \textbf{a}--\textbf{c}, Energy profiles leading to $(V_{\mathrm{Cr}}-\mathrm{Cr}_{\mathrm{i}})_{a}$ without an additional vacancy, in the presence of $V_{\mathrm{S}}^{\mathrm{in}}$, and in presence of $V_{\mathrm{Br}}^{\mathrm{in}}$.
    \textbf{d}--\textbf{f}, Corresponding energy profiles for $(V_{\mathrm{Cr}}-\mathrm{Cr}_{\mathrm{i}})_{b}$.
    Black circles and solid lines denote qNEP-NEB images and force-based interpolations, respectively.
    Orange squares in \textbf{a} and \textbf{d} show DFT results, with dashed lines.
    Energies are referenced to the initial configuration of each pathway; $\Delta E_{\mathrm{qNEP}}$ gives the final-state energy relative to this reference.
    The reaction coordinate is the cumulative atomic displacement along the path.}
    \label{fig:cr_Frenkel_pair}
\end{figure*}

For orthorhombic bilayers, DFT predicts Br vacancies to have the lowest formation energies, at 2.067--2.195~eV, followed by S vacancies at 2.822--2.855~eV and Cr vacancies at 3.186--3.194~eV.
Those nearer the surface are slightly more energetically favorable.
qNEP reproduces this ordering and the small energy differences between the in and out configurations.
Agreement also extends to the $\text{V}_{\mathrm{Cr}}$--$\text{V}_{\mathrm{Br}}$ complex and the two Cr vacancy--interstitial configurations.
In particular, DFT places the ($\text{V}_{\mathrm{Cr}}\!-\!\mathrm{Cr}_{\mathrm{i}}$)$_{b}$ configuration 1.022~eV below the ($\text{V}_{\mathrm{Cr}}\!-\!\mathrm{Cr}_{\mathrm{i}}$)$_{a}$ configuration, compared with 1.072~eV from qNEP.
The RMSE across this group is 0.101~eV, indicating that qNEP captures both the formation-energy hierarchy and the dependence on defect arrangement.

A monolayer of $m$-CrSBr comprises three atomic sheets, with a Cr plane sandwiched between two mixed SBr planes.
This structure forms when Cr atoms migrate from the interior CrS sublayer to a new crystallographic site in the vdW gap, reconstructing the bonding network into a 1T-MoS$_2$-type topology.
The migrated Cr atoms adopt distorted octahedral coordination with two S and four Br neighbors, replacing the four-S and two-Br coordination found in $o$-CrSBr; the mixed-anion outer planes and monoclinic symmetry distinguish $m$-CrSBr from the trigonal 1T-MoS$_2$ structure.

In the monoclinic phase, the Cr vacancy has the lowest DFT formation energy of 2.005~eV, whereas Br and S vacancies lie within 2.518--2.649~eV.
Thus, the lowest-energy vacancy changes from Br in the orthorhombic phase to Cr in the monoclinic phase under the adopted reference conditions, as described in Section S2.3.
qNEP reproduces this change and describes the monoclinic vacancy group with an RMSE of 0.093~eV.
The formation energies of the $\text{V}_{\mathrm{Cr}}$--$\text{V}_{\mathrm{S}}$ and $\text{V}_{\mathrm{Cr}}$--$\text{V}_{\mathrm{Br}}$ complexes are also reproduced within 0.141 and 0.003~eV, respectively.

Antisite defects present a more demanding benchmark, with RMSEs of 0.511 and 0.348~eV for the orthorhombic and monoclinic monolayers, respectively.
Nevertheless, qNEP captures the principal energetic trends.
In $o$-CrSBr, $\mathrm{Br}_{\mathrm{S}}$ is the most favorable antisite, with formation energies of 0.706~eV from DFT and 0.701~eV from qNEP.
The largest discrepancy occurs for $\mathrm{S}_{\mathrm{Br}}$, whose formation energy is underestimated by 0.936~eV.
In the monoclinic phase, DFT predicts nearly degenerate $\mathrm{Br}_{\mathrm{S}}$ and $\mathrm{S}_{\mathrm{Br}}$ defects at 0.342 and 0.345~eV. qNEP retains their low-energy character but predicts a larger separation between them.
Antisites involving Cr remain substantially more costly in both phases.
Nevertheless, the relative defect formation energies predicted by the qNEP are consistent with DFT values.
These results support the use of qNEP to investigate vacancy energetics and broad antisite trends, while identifying closely competing antisite configurations as cases requiring DFT verification.

\subsection*{Chromium Frenkel-pairs}
\label{subsec:mlvsdft}

The Cr Frenkel-pair formation provides a pathway for redistributing Cr atoms within the bilayer while conserving its overall composition.
We investigated two migration pathways leading to distinct Cr Frenkel-pairs, denoted $(V_{\mathrm{Cr}}-\mathrm{Cr}_{\mathrm{i}})_{a}$ and $(V_{\mathrm{Cr}}-\mathrm{Cr}_{\mathrm{i}})_{b}$.
Figure~\ref{fig:cr_Frenkel_pair} compares their energy profiles and examines how pre-existing S and Br vacancies modify the energetics.
The structural dynamics are shown in the Supplementary Movies, which are labeled according to the pathways in Fig.~\ref{fig:cr_Frenkel_pair}.
For each pathway, the reaction energy $\Delta E=E_{\mathrm{final}}-E_{\mathrm{initial}}$ is referenced to its corresponding initial configuration.
A $4 \times 4 \times 1$ supercell of bilayer $o$-CrSBr was used for these calculations.

In the absence of additional vacancies, qNEP predicts \emph{forward} barriers of 3.78 and 2.52~eV for Cr migration along the $a$ and $b$ axes, respectively, with reaction energies of 3.52 and 2.46~eV (\ref{fig:cr_Frenkel_pair}a and d).
The corresponding DFT profiles yield barriers of approximately 4.02 and 2.69~eV and reaction energies of 3.83 and 2.42~eV.
Thus, qNEP reproduces the energetic ordering of the two pathways and their overall profile shapes, while underestimating the forward barriers by approximately 0.24 and 0.17~eV.
The pathway to configuration $(V_{\mathrm{Cr}}-\mathrm{Cr}_{\mathrm{i}})_{a}$ exhibits multiple maxima and intervening minima, whereas that to configuration $(V_{\mathrm{Cr}}-\mathrm{Cr}_{\mathrm{i}})_{b}$ is dominated by a single maximum.

The pre-existing anion vacancies reduce both the forward barrier and the energy cost of Cr Frenkel-pair formation, with a stronger effect for S vacancies.
For $(V_{\mathrm{Cr}}-\mathrm{Cr}_{\mathrm{i}})_{a}$, an S vacancy lowers the qNEP barrier from 3.78 to 2.90~eV and the reaction energy from 3.52 to 2.48~eV.
A Br vacancy produces smaller reductions, to 3.59 and 3.37~eV, respectively.
The same trend holds for $(V_{\mathrm{Cr}}-\mathrm{Cr}_{\mathrm{i}})_{b}$: the barrier decreases from 2.52 to 1.68~eV in the presence of an S vacancy and to 2.38~eV with a Br vacancy, while the reaction energies decrease to 1.54 and 2.32~eV.
These results indicate that the local anion-vacancy environment can facilitate Cr redistribution, particularly near S vacancies.

The reverse barriers for the $(V_{\mathrm{Cr}}-\mathrm{Cr}_{\mathrm{i}})_{b}$, and in the presence of an S or Br vacancy, are 60, 136, and 65~meV, respectively.
These increased barriers indicate that vacancies, particularly S vacancies, can hinder reverse Cr migration and thereby make the migration process less reversible.
Tables~S3 and~S4 summarize the performance of the NEP variants in predicting defect formation energies.

\subsection*{Displacement threshold energy}
\label{sec:displacement}

The displacement threshold energy ($E_d$) is the minimum energy required to displace an atom from its equilibrium lattice position and create a stable defect.
Under electron irradiation, this process occurs through elastic collisions between incident electrons and atoms in the lattice.
We further evaluated the accuracy of the MLIP by calculating the displacement threshold energies of Cr, S, and Br atoms in CrSBr. The NEP-MD simulations were performed using LAMMPS ~\cite{LAMMPS} with the same $5 \times 5$ supercell of $o$-CrSBr monolayer as used in the DFT-MD calculations, ensuring a consistent comparison between the two methods.
Using the displacement threshold energies obtained from NEP-MD, we apply the McKinley--Feshbach approximation to calculate the electron-induced displacement cross sections as a function of electron beam energy~\cite{McKinley}.

\begin{table}[htbp]
    \centering
    \caption{Displacement threshold energies ($E_d$) obtained from DFT and NEP for different atomic sites in CrSBr.
    The superscripts ``in'' and ``out'' denote atomic sites located closer to the inner region and the outer surface of the monolayer, respectively.}
    \label{tab:displacement_thresholds}
    \begin{tabular}{lcccc}
        \hline
        Method & Br & Cr$^{\text{out}}$ & Cr$^{\text{in}}$ & S \\
        \hline
        DFT     & 3.45 & 13.55 & 10.30 & 6.70 \\
        NEP & 3.90 & 13.70 & 11.80 & 7.00 \\
        \hline
    \end{tabular}
\end{table}
Table~\ref{tab:displacement_thresholds} compares the displacement threshold energies obtained from DFT and NEP for different atomic sites. Overall, NEP reproduces the DFT results reasonably well for Br, Cr$^{\text{out}}$, and S, with differences of about 0.5~eV or less. A larger deviation is observed for the Cr$^{\text{in}}$, where NEP predicts a threshold of 11.80~eV compared with 10.30~eV from DFT. Both methods predict Br to have the lowest displacement threshold, indicating that it is the most susceptible to displacement under electron irradiation.

\begin{figure}[!t]
    \centering
    \includegraphics[width=0.45\textwidth]{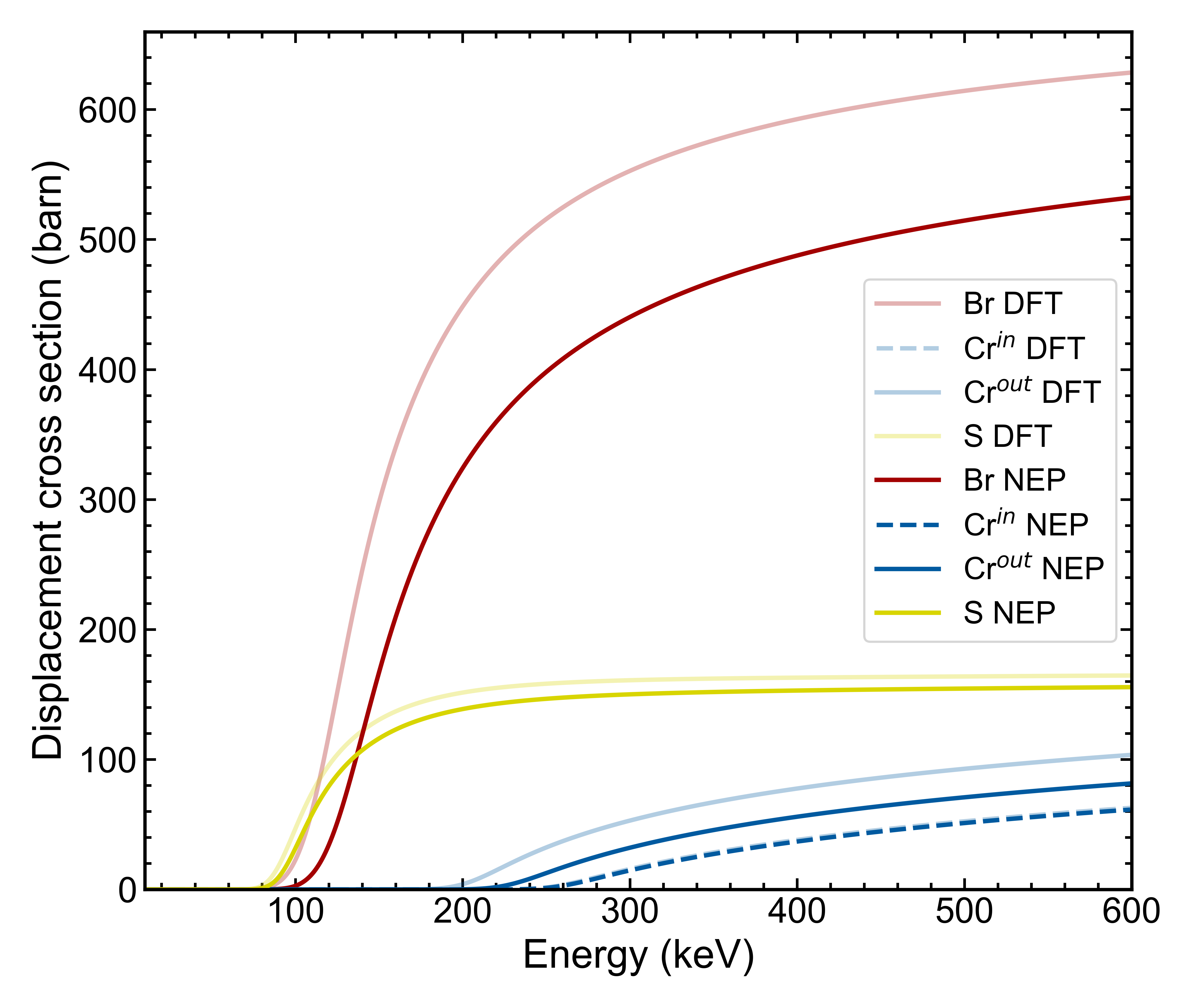}
    \caption{\textbf{Displacement cross sections} for Cr, S and Br atoms in CrSBr plotted as functions of electron energy, calculated using DFT-MD and NEP-MD combined with the McKinley–Feshbach formalism.}
    \label{fig:displacement_threshold}
\end{figure}

As show in Fig.~\ref{fig:displacement_threshold}, our theoretical results indicate that Cr atoms in CrSBr can be displaced by 300~keV electrons, consistent with recent experimental observations of electron-beam-induced Cr displacement and the formation of vacancy--interstitial complexes~\cite{Klein2026}.
The displacement of Cr atoms is particularly important for understanding the beam-induced structural rearrangement and phase transformation in $o$-CrSBr, as their migration into the vdW gap can lead to the formation of new structural phases.~\cite{klein2022control}

\section*{Discussion}
\label{sec:discussion}

The MLIPs developed here achieve near-DFT accuracy across a chemically and structurally diverse configuration space, establishing a practical computational framework for CrSBr that extends well beyond the capabilities of first-principles methods alone.
Among the three NEP variants, qNEP provides the most balanced description: it reproduces lattice constants, formation enthalpies, and the strongly anisotropic pressure response of $o$-CrSBr, while offering the most reliable account of the Cr--Cr interaction at short range.
For defect energetics, qNEP captures the formation-energy hierarchy of vacancies and vacancy complexes with RMSEs below 0.10 eV in both phases, sufficient to rank competing defects reliably, though antisite defects remain more challenging, with RMSEs up to 0.5 eV, identifying them as configurations where DFT verification is warranted.
The NEB calculations further reveal that qNEP systematically underestimates Cr Frenkel-pair barriers relative to DFT, consistent with the known tendency of NEP-type models to soften transition-state energies; this should be treated as a lower bound on migration barriers when interpreting qNEP-driven kinetic results.
Crucially, the finding that pre-existing S vacancies reduce the Cr Frenkel-pair barrier by up to 1.1 eV, far more than Br vacancies, provides an atomistic rationale for the experimentally observed irradiation sensitivity of CrSBr: the high native concentration of bromine vacancies generates a defect landscape in which subsequent S vacancy formation opens preferential channels for Cr redistribution, linking the stochastic defect inventory to the deterministic onset of structural transformation.
The calculated displacement threshold energies corroborate this picture: the markedly low threshold of Br ($E_d \approx 3.5$ eV) confirms that anion vacancies are the primary radiation-induced defects at accessible beam energies, priming the lattice for the Cr redistribution pathways identified above.

Beyond the immediate benchmarks, these results carry broader implications for defect engineering and phase-transformation control in vdW magnets.
The reversal of vacancy stability, from V$_\text{Br}$ being lowest in energy in $o$-CrSBr to V$_\text{Cr}$ being lowest in $m$-CrSBr, suggests that the two polymorphs sustain qualitatively different intrinsic defect populations under equilibrium conditions, which in turn will drive distinct doping landscapes and magnetic perturbations.
This phase-dependent defect chemistry, now accessible at large scale through qNEP, opens a route to deliberately biasing the material towards one defect regime or the other through choice of growth conditions or post-growth treatment.
Looking forward, the present framework can be extended in several directions: incorporating spin degrees of freedom via spin-lattice potentials~\cite{Yu2024_APS} would resolve the magnetic-state ambiguity that currently limits the description of magnon-defect coupling and finite-temperature magnetic order; coupling qNE--driven MD with Monte Carlo sampling of chemical disorder would enable quantitative defect-concentration phase diagrams; and applying the displacement threshold energies reported here to realistic beam-current models would allow direct prediction of defect-creation yields as a function of electron-beam voltage, supporting the rational design of irradiation protocols for site-selective phase patterning in CrSBr-based magneto-optoelectronic devices.

\section*{Acknowledgments}
\label{sect:ack}
We are grateful to  CSC--IT Center for Science Ltd. and the Aalto Science-IT project for generous grants of computer time.
X. C., and T.A-N. acknowledge support from the Research Council of Finland grant no. 370057, and A. H. and M. K. were supported by the Research Council of Finland Flagship of Advanced Mathematics for Sensing Imaging and Modelling, grant 358944 and by Foundation PS. 
Generous CPU time grants from the Paderborn Center for Parallel Computing (PC2, Noctua 2 cluster, hpc-prf-def2dhet) and Gauss Centre for Supercomputing (GCS), Supercomputer HAWK at Höchstleistungsrechenzentrum Stuttgart, are greatly appreciated.

\section*{Data Availability}
The code used for dataset construction, machine-learning interatomic potential training, and data analysis, together with the trained potential models, will be made publicly available on GitHub upon acceptance of the manuscript.

\bibliography{main.bib}
\end{document}